\documentclass[twocolumn,preprintnumbers,amsmath,amssymb,superscriptaddress]{revtex4-2}

\usepackage{float}
 \usepackage{booktabs}
\usepackage[caption=false]{subfig}
\usepackage{amsmath}
\usepackage{comment}
\UseRawInputEncoding

\usepackage{latexsym,amssymb,amsthm,amsmath,epsfig,braket}
\usepackage[left=2.00cm, right=2.00cm, top=2.00cm, bottom=2.00cm]{geometry}
\usepackage{xcolor}
\usepackage{svg}
\usepackage[colorlinks, citecolor={blue!80!black}, urlcolor={blue!80!black}, linkcolor={red!50!black}]{hyperref}
\usepackage{float}
\usepackage{hyperref}
\hypersetup{
    citecolor=red,
    colorlinks=true,
    linkcolor=blue,
    filecolor=blue,      
    urlcolor=blue,
}

\begin{document}

\title{Interlayer Coupling and Floquet-Driven Topological Phases in Bilayer Haldane Lattices}

\author{Imtiaz Khan}
\affiliation{Department of Physics, Zhejiang Normal University, Jinhua 321004, P. R. China}
\affiliation{Research Center of Astrophysics and Cosmology, Khazar University, Baku, AZ1096, 41 Mehseti Street, Azerbaijan}
\affiliation{ Zhejiang Institute of Photoelectronics $\&$ Zhejiang Institute for Advanced Light Source, Zhejiang Normal University, Jinhua, 321004, P. R. China}
\author{Muzamil Shah}
\email{muzamil@qau.edu.pk}
\affiliation{Department of Physics, Quaid-I Azam University Islamabad, 45320, Pakistan}
\affiliation{Research Center of Astrophysics and Cosmology, Khazar University, Baku, AZ1096, 41 Mehseti Street, Azerbaijan}
\author{Reza Asgari}
\affiliation{Department of Physics, Zhejiang Normal University, Jinhua 321004, P. R. China}
\affiliation{School of Physics, Institute for Research in Fundamental Sciences (IPM), Tehran 19395-5531, Iran}

\author{Gao Xianlong}
\email{gaoxl@zjnu.edu.cn}
\affiliation{Department of Physics, Zhejiang Normal University, Jinhua 321004, P. R. China}


\begin{abstract}
We investigate Floquet-driven topological phase transitions in an AB-stacked bilayer Haldane lattice with tunable intralayer hopping anisotropy. By combining interlayer hybridization, Haldane flux, and off-resonant circularly polarized light, we obtain controlled transitions among Dirac, semi-Dirac, and higher-Chern insulating phases. As the hopping anisotropy increases, the two inequivalent Dirac points move toward each other and merge at the Brillouin-zone $\mathbf{M}$ point, where a semi-Dirac dispersion emerges with linear and quadratic momentum dependence along orthogonal directions. In this regime, competition between the intrinsic Haldane mass and the Floquet-induced mass drives a sequence of sharp topological transitions with Chern numbers $C=0,\pm1,\pm2$. We further show that interlayer coupling qualitatively reshapes the Floquet band topology by inducing helicity-dependent and valley-selective band inversions at the K and K$'$ points, thereby stabilizing higher-Chern phases in the valence bands. These changes are accompanied by redistribution of the Berry curvature, bulk gap closings, and the collapse or sign reversal of quantized anomalous Hall plateaus. As the system approaches the semi-Dirac limit, the topological phase space narrows and disappears at the critical merger point, beyond which the system becomes topologically trivial even when it remains gapped. Overall, the bilayer geometry broadens the scope of Floquet topological control by enabling dynamically tunable higher-Chern phases and valley-dependent Hall responses governed by interlayer coupling and light helicity.

\end{abstract}

\maketitle
\newpage
\section{Introduction}
The Haldane model \cite{Haldane1988} is a foundational example of a topological system. It showed that the quantum Hall effect can arise in a two-dimensional honeycomb lattice without an external magnetic field. This behavior originates from broken time-reversal symmetry introduced through complex next-nearest-neighbor hopping, which gives the bands a nonzero Chern number and produces a Chern-insulating phase. Although the original model describes a single honeycomb layer, attention has increasingly shifted toward layered Dirac systems. Compared with monolayers, bilayer and multilayer structures support quadratic band dispersions and can develop a finite gap at the Dirac points set by the interlayer tunneling strength. These features enrich the underlying physics and create new opportunities for topological design. As a result, extensions of the Haldane model have been explored in coupled two-dimensional systems, including bilayer materials \cite{Spurrier2020, ChengP2019, SornS2018, PanasJ2020} and moir\'e superlattices \cite{XiaoY2020,YouYZ2019,TongQ2017,LauCN2022,AndreiEY2021}.

Alongside these developments, band-structure engineering has been widely studied in systems such as monolayer graphene \cite{MondalS2021}, spin Hall insulators \cite{MondalSbs2022}, and dice lattices \cite{MondalSbss2023}. One simple and effective method is to introduce anisotropy into the nearest-neighbor hopping amplitudes of the honeycomb lattice. In practice, one hopping parameter along a chosen bond direction, denoted $t_1$, is tuned independently while the other two remain equal to $t$, with $t\sim 3$ eV in graphene. As $t_1$ increases, the two Dirac cones move toward each other and merge at the $\mathbf{M}$ point of the Brillouin zone when $t_1 = 2.0t$. At this critical value, known as the \emph{semi-Dirac point}, the band gap closes and the system loses its nontrivial topology. The resulting dispersion is linear along $k_x$ and quadratic along $k_y$, which is the defining feature of semi-Dirac behavior. Such dispersions have been reported in several experimental settings, including TiO$_2$/VO$_2$ multilayer heterostructures \cite{PardoV2009,PardoVp2010}, doped and pressurized monolayer phosphorene \cite{RodinAS2014,GuanJ2014}, organic salts such as BEDT--TTF$_2$I$_3$ under pressure \cite{SuzumuraY2013,HasegawaY2006}, and black phosphorus with potassium surface doping \cite{ZhongC2017}. A related route is uniaxial strain, which selectively changes bond lengths and angles, and therefore the hopping amplitudes, along the strain direction while leaving the others largely unchanged. This approach has also been shown to generate semi-Dirac features in monolayer honeycomb materials such as Si$_2$O \cite{ZhongC2017}.

While these ideas have been studied extensively in monolayer systems, their role in multilayer structures, especially bilayer graphene, remains much less explored. Bilayers offer a broader parameter space and therefore a richer phase diagram, making them a promising setting in which to search for new topological behavior. In particular, edge states and quantized Hall transport in engineered bilayer systems are of strong interest. The additional bands present in the bilayer spectrum also allow higher Chern numbers to emerge. These phases are important because they increase the anomalous Hall response and support multiple chiral edge channels in a semi-infinite geometry, leading to richer transport signatures. Higher-Chern phases have been identified in a wide range of settings, including Dirac \cite{SticletD2013} and semi-Dirac systems \cite{SSMondal2022} with extended hopping, multiorbital triangular lattices \cite{YangSGu2012}, decorated honeycomb lattices \cite{ChenWC2012}, honeycomb systems with spin-orbit coupling \cite{YangYZhang2014,YangYli2016}, and ultracold atoms in triangular optical lattices \cite{AlaseA2021,ŁąckiM2021}. Related behavior has also been reported in solid-state systems such as magnetically doped topological insulators \cite{FangCG2014}, Cr-doped Bi$_2$(Se,Te)$_3$ thin films \cite{WangJlian2013}, and MnBi$_2$Te$_4$ at elevated temperatures \cite{ZhuWSong2022,GeJLiu2020}, as well as in engineered multilayer heterostructures \cite{ZhaoYF2020} and classical platforms including acoustic metamaterials and sonic crystals \cite{ZhaoHZhang2022}.

A complementary route to controlling band topology is provided by periodic driving. According to Floquet theory \cite{Sambe:1973cnm, Shirley:1965rgd}, the dynamics of a system under high-frequency driving can often be described by an effective time-independent Hamiltonian that captures the slow evolution. Exact expressions for this Hamiltonian are rarely available, so one typically relies on approximations such as the Magnus expansion~\cite{Magnus:1954zz, Blanes2009MagnusExpansion} or other high-frequency expansions~\cite{Maricq1982hfeFloquetNmrOfSolids,Rahav2003HighFreqExp1}. These methods are perturbative, and their convergence is not always guaranteed, even in noninteracting systems. The problem becomes more subtle in the presence of interactions, where heating can limit the validity of the expansion~\cite{Prosen1999,Prosen2011,Ponte2015,banin2015,D’Alessio,Mori2016}. Even so, Floquet engineering has become a powerful framework for designing effective Hamiltonians with desired properties through time-periodic driving \cite{Goldmann2014,Eckardthf2015,BukovMd2015}. Using this approach, several studies have predicted Floquet topological insulators in irradiated graphene-based systems \cite{KitagawaT2011,OkaT2009,Gómez-León2014}, with experimental confirmation reported in driven materials \cite{WangYH2013}. Periodically driven superconductors have likewise been proposed as candidates for realizing Floquet Majorana modes \cite{Jianglkt,TongQ.J,WangZB}.

Despite this progress, two challenges continue to limit the broader realization of Floquet engineering. First, it is not enough to reshape the quasienergy spectrum; the relevant Floquet bands must also be occupied appropriately. Observable quantities such as the Chern number depend on the Berry curvature of the occupied states, so band occupation is essential. Second, many studies of driven systems have focused mainly on replica bands and hybridization gaps \cite{WangYH2013,MahmoodF2016}, rather than on achieving deeper and more selective control over band topology itself. These limitations motivate the search for platforms in which periodic driving can be combined with structural degrees of freedom to produce a wider range of robust topological phases.

In this work, we study a bilayer graphene system with broken time-reversal symmetry, modeled as a coupled bilayer Haldane lattice in conventional Bernal (AB) stacking, where the $B$ sublattice of the top layer lies directly above the $A$ sublattice of the bottom layer. This system already exhibits a rich topological band structure: some bands carry Chern numbers $\pm2$ and $\pm1$, while others carry only $\pm1$, reflecting the combined effects of interlayer hybridization and broken symmetry. This structure makes the bilayer setting especially well suited for controlled topological band engineering. In particular, interlayer coupling naturally supports $|C|=2$ sectors, allowing fuller sequences of Floquet-driven phase transitions and multichannel topological transport that are not available in monolayer systems. Motivated by this, we examine how asymmetric nearest-neighbor hopping within each layer, with fixed interlayer tunneling, can be used to drive topological transitions. We also include off-resonant circularly polarized light as an external control field, which generates an effective Floquet Haldane mass and enables dynamic tuning of both the band gaps and the Berry curvature. Through this bilayer Floquet-Haldane framework, we show how hopping anisotropy, interlayer coupling, and light helicity together determine the appearance, evolution, and eventual collapse of distinct topological phases.

\section{The Hamiltonian and the introduction to the Floquet formalism}\label{sec:model_hamiltonian}

\begin{widetext}
A tight-binding description of the bilayer honeycomb lattice can be expressed as \cite{Mondal2023}:

\begin{figure}[h]
    \centering \includegraphics[width=12.90cm]{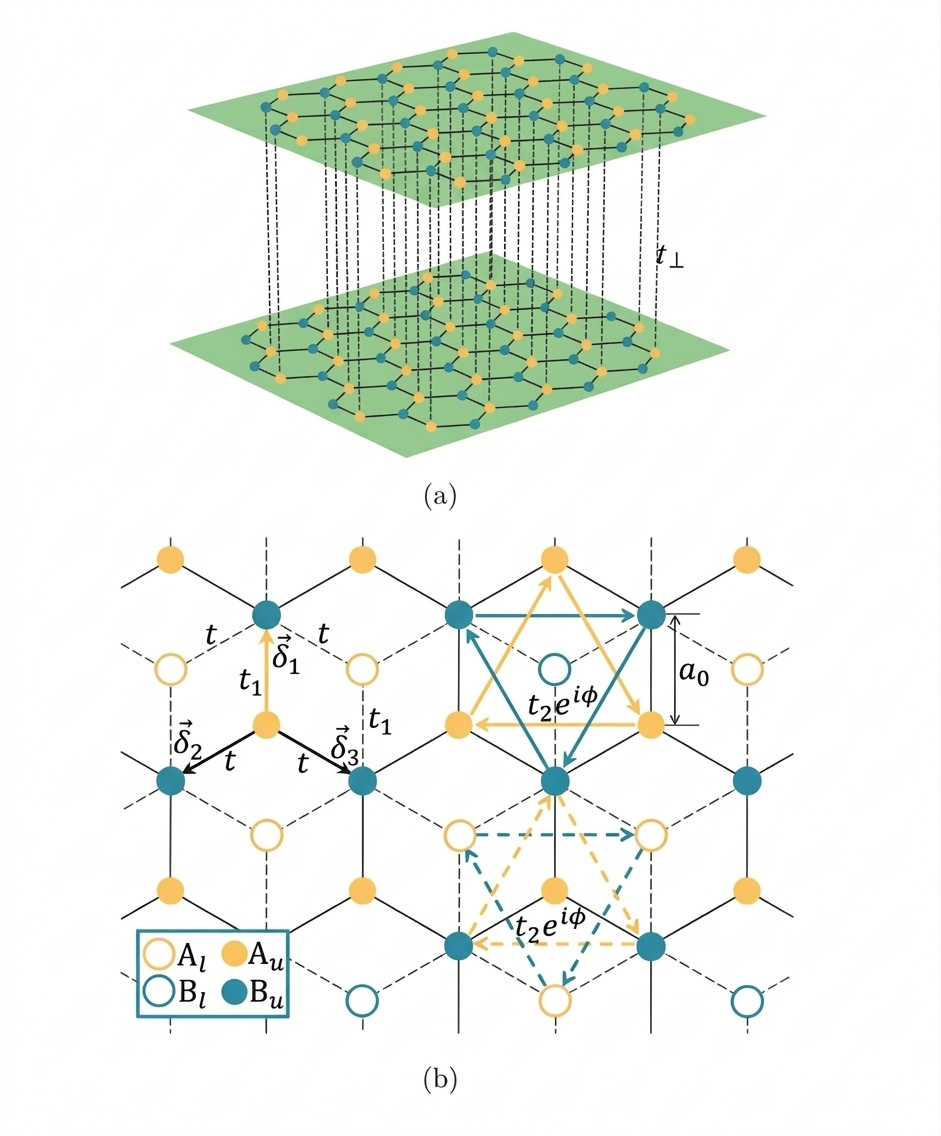}
    \caption{\raggedright 
    (a) Schematic of AB-stacked bilayer graphene, where the interlayer hopping $t_\perp$ couples 
    the $B$ sublattice of the upper layer to the $A$ sublattice of the lower layer. 
    In both layers, $A$ and $B$ sites are represented by red and blue circles, respectively. 
    (b) Illustration of the in-plane hopping processes. To clearly distinguish the two layers, 
    the sublattices of the lower sheet are denoted as $\mathrm{A}_l$ (yellow) and $\mathrm{B}_l$ (green), 
    while those of the upper sheet are labeled $\mathrm{A}_u$ and $\mathrm{B}_u$. 
    Nearest-neighbor bonds within the lower layer are shown as dashed lines, and the corresponding 
    next-nearest-neighbor hoppings as dashed arrows. The hopping amplitude along the 
    $\boldsymbol{\delta}_1$ direction (highlighted by the yellow arrow) is $t_1$, 
    while the hoppings along $\boldsymbol{\delta}_{2,3}$ are $t$ 
    (with $\boldsymbol{\delta}_i$ defined in the text). 
    The next-nearest-neighbor processes carry complex amplitudes 
    $t_2 e^{i\phi}$ for clockwise paths and $t_2 e^{-i\phi}$ for anticlockwise paths.
    }
    \label{mainfig}
\end{figure}
    
\begin{figure}[h!]
	\centering \includegraphics[width=15.90cm]{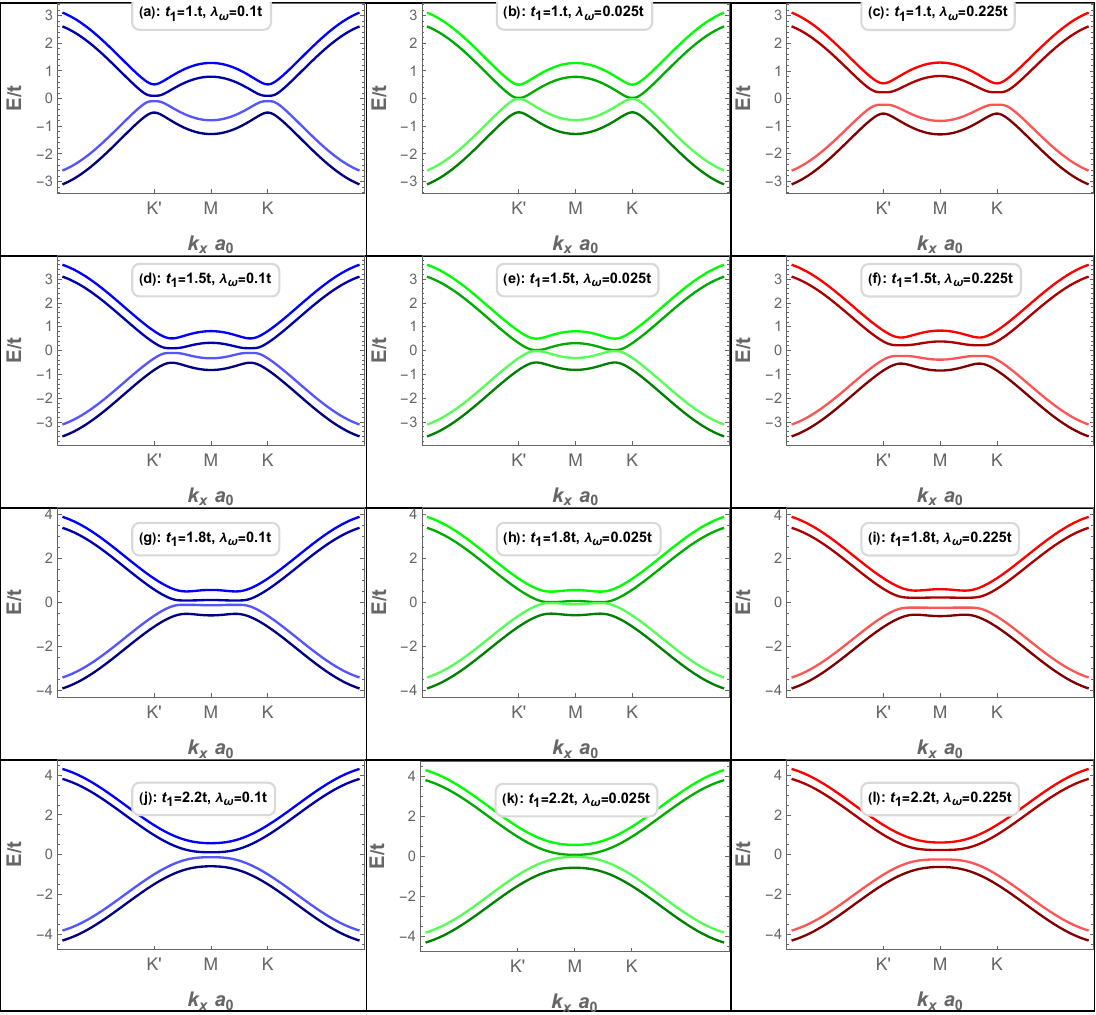}
 \caption{Band structures along the $k_x a_0$ direction at fixed $k_y a_0 = 2\pi/3$. 
Panels (a)--(l) correspond to finite next-nearest-neighbor hopping $t_2 = 0.0t$ with 
$t_1 = t$, $1.5t$, $1.8t$, and $2.2t$, respectively. For comparison, panels (e)--(h) 
show the dispersions in the absence of $t_2$ for the same sequence of 
$t_1$ values. The remaining parameters are fixed at $t_\perp = 0.5t$,  
and $\phi_l = \phi_u = \pi/2$.}
		\label{f1}
\end{figure}
\begin{figure}[h!]
	\centering \includegraphics[width=15.90cm]{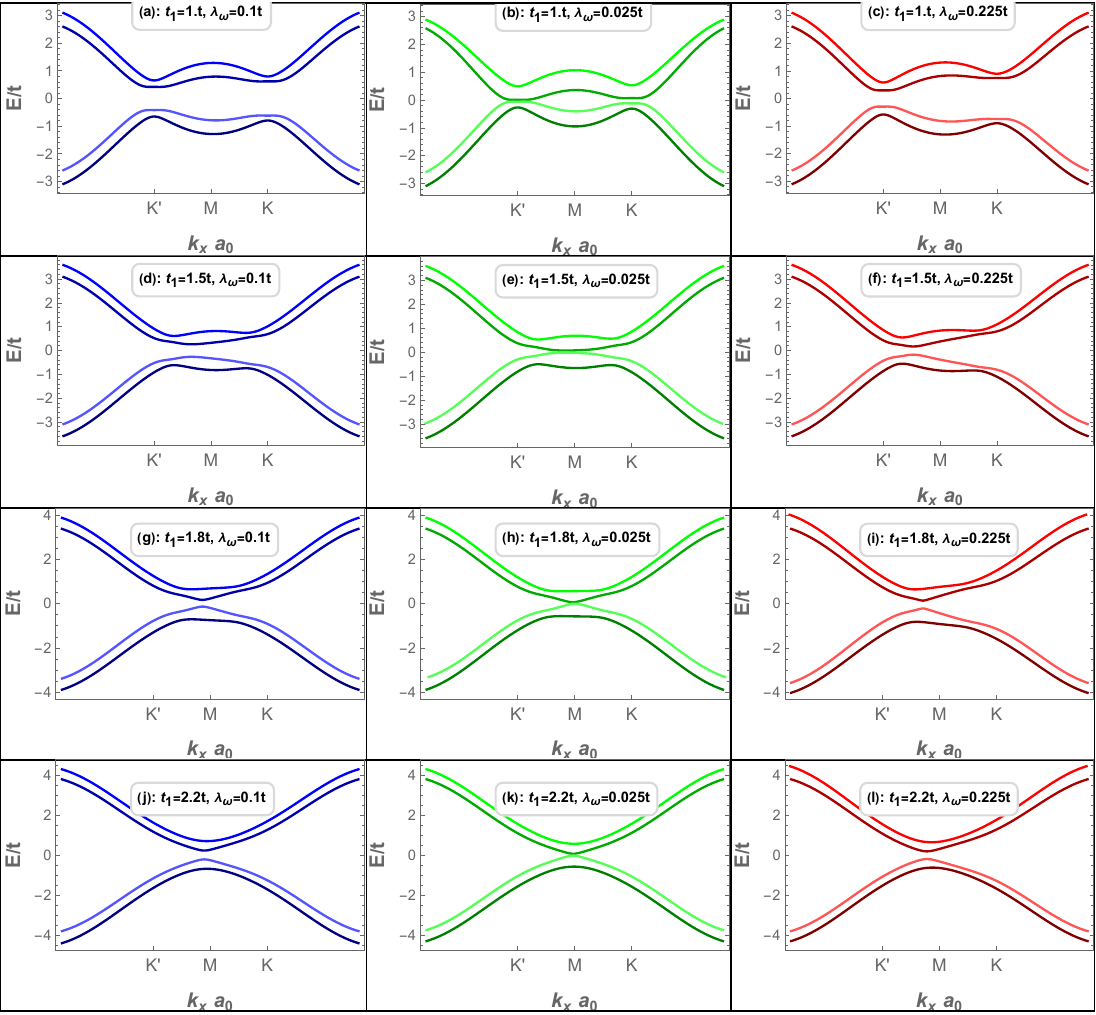}
   \caption{Band structures along the $k_x a_0$ direction at fixed $k_y a_0 = 2\pi/3$. 
Panels (a)--(l) correspond to finite next-nearest-neighbor hopping $t_2 = 0.1t$ with 
$t_1 = t$, $1.5t$, $1.8t$, and $2.2t$, respectively. For comparison, panels (e)--(h) 
show the dispersions in the presence of $t_2$, the same sequence of 
$t_1$ values. The remaining parameters are fixed at $t_\perp = 0.5t$,  
and $\phi_l = \phi_u = \pi/2$.}
		\label{f2}
\end{figure}

\begin{figure*}[h]
\centering
\subfloat[Berry curvature distributions of the valence band $v_1$ for $\lambda_\omega = 0.1t$.%
\label{f7a}]{%
\includegraphics[width=0.78\textwidth]{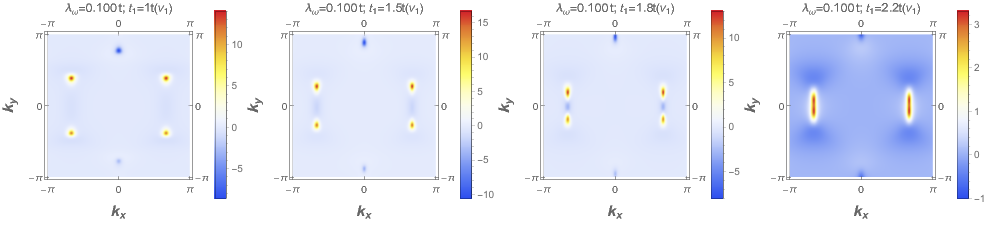}}
\hfill
\subfloat[Berry curvature distributions of the valence band $v_1$ for $\lambda_\omega = 0.025t$.%
\label{f7b}]{%
\includegraphics[width=0.78\textwidth]{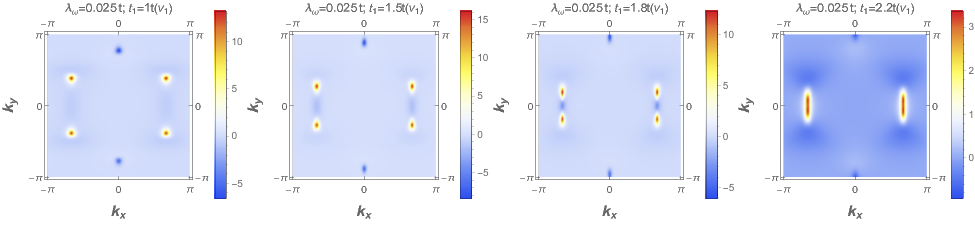}}\\[0.8ex]
\subfloat[Berry curvature distributions of the valence band $v_1$ for $\lambda_\omega = 0.225t$.%
\label{f7c}]{%
\includegraphics[width=0.78\textwidth]{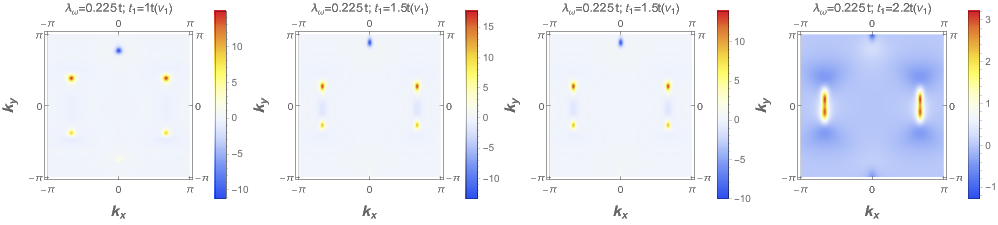}}
\caption{Berry curvature distributions of the valence band $v_1$ for 
$t_1 = t$, $t_1 = 1.5t$, $t_1 = 1.8t$, and $t_1 = 2.2t$. The other parameters are fixed as 
$\phi_l = \phi_u = \pi/2$, and $t_\perp = 0.5t$,.}
\label{f7}
\end{figure*}

\begin{figure*}[h]
\centering
\subfloat[Berry curvature distributions of the valence band $v_2$ for $\lambda_\omega = 0.1t$.%
\label{f8a}]{%
\includegraphics[width=0.78\textwidth]{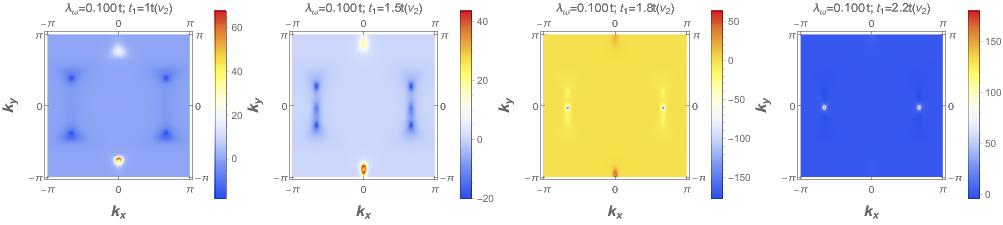}}
\hfill
\subfloat[Berry curvature distributions of the valence band $v_2$ for $\lambda_\omega = 0.025t$.%
\label{f8b}]{%
\includegraphics[width=0.78\textwidth]{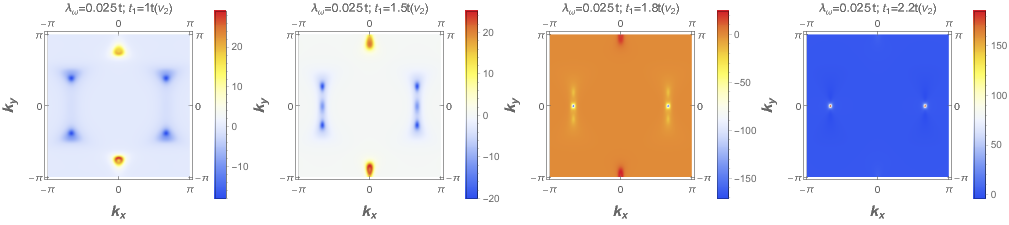}}\\[0.8ex]
\subfloat[Berry curvature distributions of the valence band $v_2$ for $\lambda_\omega = 0.225t$.%
\label{f8c}]{%
\includegraphics[width=0.78\textwidth]{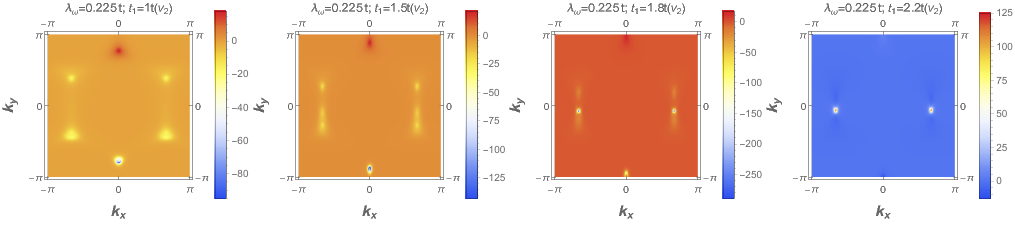}}
\caption{Berry curvature distributions of the valence band $v_2$ for 
$t_1 = t$, $t_1 = 1.5t$, $t_1 = 1.8t$, and $t_1 = 2.2t$. The other parameters are fixed as 
$\phi_l = \phi_u = \pi/2$, and $t_\perp = 0.5t$,.}
\label{f8}
\end{figure*}

\begin{figure*}[h]
\centering
\subfloat[Phase diagram of the lowest occupied band ($v_1$) for 
$t_\perp = 0.5t$ with $\lambda_\omega = 0.1t$.%
\label{f9a}]{%
\includegraphics[width=0.78\textwidth]{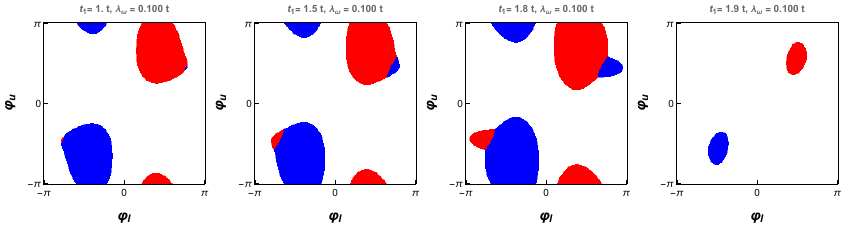}}
\hfill
\subfloat[Phase diagram of the lowest occupied band ($v_1$) for 
$t_\perp = 0.5t$ with $\lambda_\omega = 0.025t$.%
\label{f9b}]{%
\includegraphics[width=0.78\textwidth]{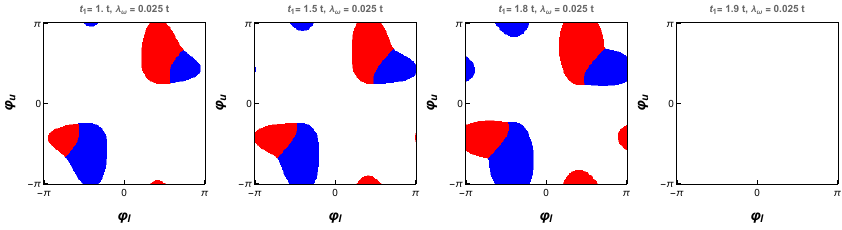}}\\[0.8ex]
\subfloat[Phase diagram of the lowest occupied band ($v_1$) for 
$t_\perp = 0.5t$ with $\lambda_\omega = 0.225t$.%
\label{f9c}]{%
\includegraphics[width=0.78\textwidth]{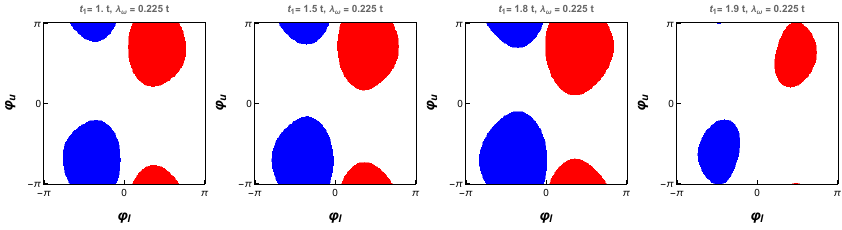}}
\caption{
White regions correspond to the trivial phase 
with Chern number $C=0$, while colored regions indicate 
topologically nontrivial phases. The color scheme is as follows: 
white ($C=0$), blue ($C=-1$), and red ($C=1$).}
\label{f9}
\end{figure*}

\begin{figure*}[t]
\centering
\subfloat[Phase diagram of the upper occupied band ($v_2$) for 
$t_\perp = 0.5t$ with $\lambda_\omega = 0.1t$.%
\label{f10a}]{%
\includegraphics[width=0.78\textwidth]{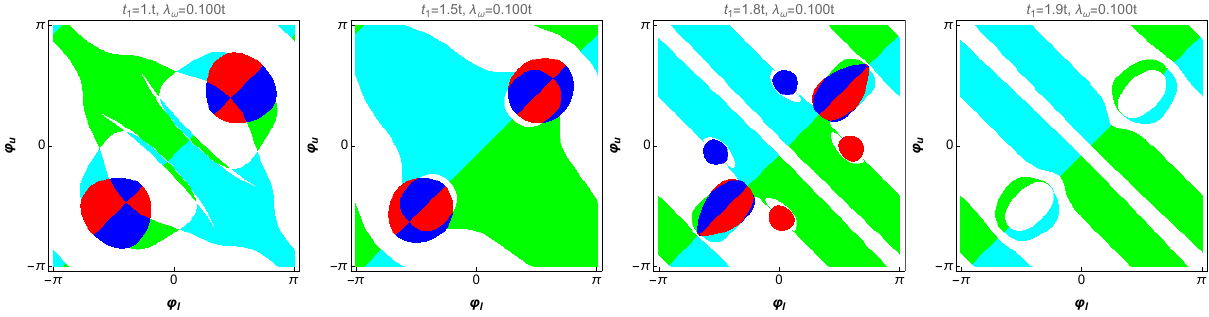}}
\hfill
\subfloat[Phase diagram of the upper occupied band ($v_2$) for 
$t_\perp = 0.5t$ with $\lambda_\omega = 0.025t$.%
\label{f10b}]{%
\includegraphics[width=0.78\textwidth]{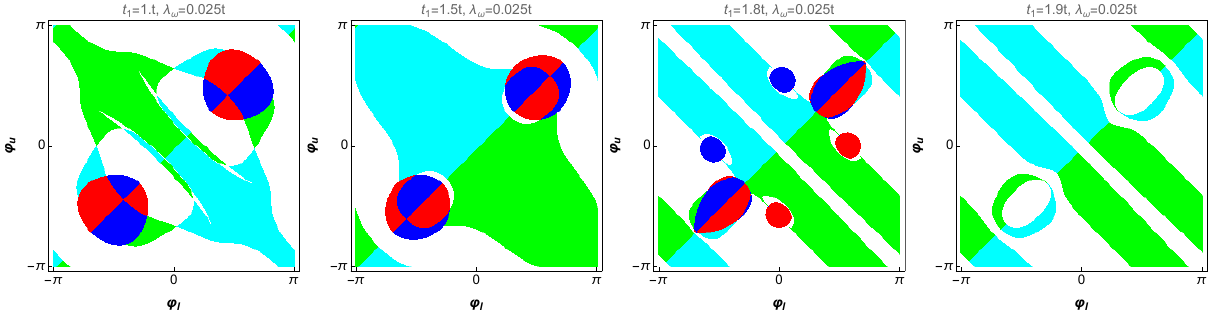}}\\[0.8ex]
\subfloat[Phase diagram of the upper occupied band ($v_2$) for 
$t_\perp = 0.5t$ with $\lambda_\omega = 0.225t$.%
\label{f10c}]{%
\includegraphics[width=0.78\textwidth]{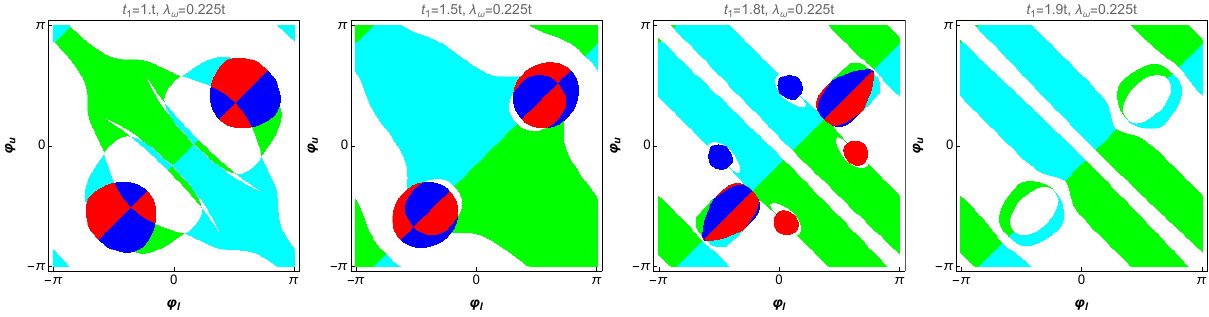}}
\caption{
White regions correspond to the trivial phase with Chern number $C=0$, while colored regions indicate 
topologically nontrivial phases. The color scheme is as follows: 
white ($C=0$), blue ($C=-1$), red ($C=1$), cyan ($C=2$) and green ($C=-2$).}
\label{f10}
\end{figure*}

	\begin{figure}[h!]
	\centering \includegraphics[width=7.90cm]{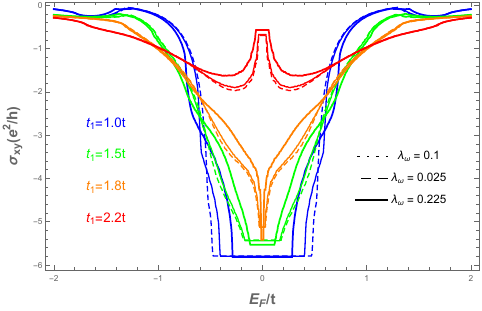}
	\centering \includegraphics[width=7.90cm]{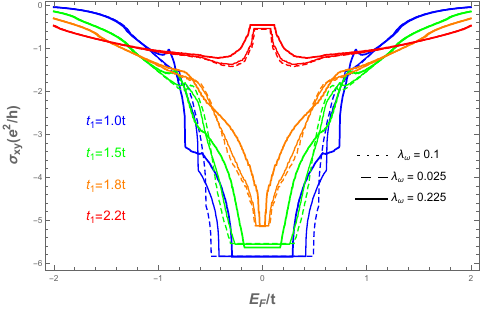}
\caption{%
Anomalous Hall conductivity as a function of the Fermi energy $E_F$ for different values of $t_1$. 
Left and right panel correspond to interlayer couplings $t_{\perp}=1.5t$ and $t_{\perp}=0.5t$, respectively. 
One clearly observes that the plateau region narrows progressively as $t_1$ departs from $t$, 
highlighting the sensitivity of the quantized response to the hopping asymmetry. The other parameters are fixed as  $t_1 = t$, $t_1 = 1.5t$, $t_1 = 1.8t$, and $t_1 = 2.2t$, 
with $t_2 = 0.1t$. 
Results are shown for different Floquet driving frequencies: 
$\lambda_\omega = 0.025$ (dotted line), $\lambda_\omega = 0.1$ (dash line), 
and $\lambda_\omega = 0.225$ (solid line).}\label{fhc}
\end{figure}

\end{widetext}

\begin{align}\label{eq:ham1}
	H = & \sum_{p\in \{l, u\}} \Bigg[ \sum_{\langle ij \rangle} t_{ij}\, c_i^{p\dagger} c^p_j 
	+ t_2 \sum_{\langle\langle i m \rangle\rangle} e^{i\phi^{im}_p} c_i^{p\dagger} c^p_m 
	+ \mathrm{h.c.} \Bigg] \nonumber\\ 
	& + \Bigg[ t_\perp \sum_{\langle q, r \rangle_\perp} c_q^{l\dagger} c^u_r + \mathrm{h.c.} \Bigg],
\end{align}
where $c_i^{p\dagger}$ ($c_i^p$) creates (annihilates) an electron at site $i$ in layer $p$, with $p=l, u$ denoting the lower and upper sheets, respectively.  

The first term on the right-hand side corresponds to nearest-neighbor (NN) hopping. The amplitude $t_{ij}$ equals $t_1$ if the bond lies along $\boldsymbol{\delta}_1 = a_0 (0,1)$, and equals $t$ (which is )in 1 to 3 eV if it lies along $\boldsymbol{\delta}_2 = a_0 (\sqrt{3}/2, -1/2)$ or $\boldsymbol{\delta}_3 = a_0 (\sqrt{3}/2, 1/2)$, as depicted in Fig.~\ref{mainfig}. The lattice constant is $a_0$. The second term incorporates complex next-nearest-neighbor (NNN) hopping with magnitude $t_2$ and a phase $\phi^{im}_p$. The flux phases, which represent the Peierls substitution implemented in the lattice model, associated with the lower and upper layers are denoted as $\phi^{im}_l$ and $\phi^{im}_u$, respectively: hopping counterclockwise contributes a positive phase, while clockwise hopping contributes a negative one. The final term describes interlayer tunneling of strength $t_\perp$, which connects the B-sublattice of the upper layer ($r \in \mathrm{B}_u$) with the A-sublattice of the lower layer ($q \in \mathrm{A}_l$), consistent with AB (Bernal) stacking.  

In our analysis, we vary $t_1$ in both layers from $t$ up to $2t$ (the semi-Dirac limit), and also explore the regime $t_1 > 2t$. Notice that by varying $t_1$, the value of $t_{\perp}$ may slightly alter; we neglect the variation of $t_{\perp}$. Such anisotropy can arise from, for instance, uniaxial strain, lattice distortion, or engineered hopping in artificial honeycomb lattices.

Upon Fourier transformation, the Hamiltonian is conveniently written in the four-sublattice basis  
$\{ \mathrm{A}_l, \mathrm{B}_l, \mathrm{A}_u, \mathrm{B}_u \}$ as
\begin{equation}\label{eq:ham_kspace}
	H(\mathbf{k}) = 
	\begin{pmatrix}
		h^+_z(\mathbf{k}, \phi_l) & h_{xy}(\mathbf{k}, t_1) & 0 & t_\perp \\
		h_{xy}^*(\mathbf{k},  t_1) & h_z^-(\mathbf{k}, \phi_l) & 0 & 0 \\
		0 & 0 & h^+_z(\mathbf{k}, \phi_u) & h_{xy}(\mathbf{k}, t_1) \\
		t_\perp & 0 & h_{xy}^*(\mathbf{k}, t_1) & h^-_z(\mathbf{k}, \phi_u) \\
	\end{pmatrix}.
\end{equation}
Here, $h_z^\pm$ are defined as  
$h^+_z(\mathbf{k}, \phi_p) = h_0(\mathbf{k}, \phi_p) + h_z(\mathbf{k}, \phi_p)$  
and  
$h^-_z(\mathbf{k}, \phi_p) = h_0(\mathbf{k}, \phi_p) - h_z(\mathbf{k}, \phi_p)$.  
The off-diagonal term $h_{xy}(\mathbf{k}, t_1)$ takes the form  
$h_{xy}(\mathbf{k}, t_1) = h_x(\mathbf{k}, t_1) - i h_y(\mathbf{k}, t_1)$.  
The explicit expressions are
\begin{align}
	h_0(\mathbf{k}, \phi_p) &= 2t_2\cos \phi_p \left[ 2\cos\!\left(\tfrac{\sqrt{3}s_x}{2}\right)\cos\!\left(\tfrac{3s_y}{2}\right) + \cos(\sqrt{3}s_x) \right], \\
	h_z(\mathbf{k}, \phi_p) &= -2t_2\sin \phi_p \left[ 2\sin\!\left(\tfrac{\sqrt{3}s_x}{2}\right)\cos\!\left(\tfrac{3s_y}{2}\right) - \sin(\sqrt{3}s_x) \right], \\
	h_x(\mathbf{k}, t_1) &= t_1 \cos s_y + 2t \cos \tfrac{s_y}{2}  \cos \tfrac{\sqrt{3}s_x}{2}, \\
	h_y(\mathbf{k}, t_1) &= -t_1 \sin s_y + 2t \sin \tfrac{s_y}{2} \cos \tfrac{\sqrt{3}s_x}{2}.
\end{align}
where $\mathbf{s}=\mathbf{k}a_0$.
Throughout this work, we set the NNN hopping amplitude to $t_2 =0.0, \, 0.1t$, and examine two representative choices of interlayer coupling, $t_\perp = 0.5t$ and $t_\perp = 1.5t$. The fluxes are chosen as $\phi_l = \phi_u = \pi/2$.

For this configuration, the band dispersions are given by
\begin{align}\label{eq:E_k_conduction}
	E^c_\pm &= h_0 + \sqrt{\tfrac{t_\perp^2}{2} + |h_{xy}|^2 + h_z^2 
	\pm \tfrac{t_\perp}{2}\sqrt{t_\perp^2 + 4|h_{xy}|^2}}, \\
	E^v_\pm &= h_0 - \sqrt{\tfrac{t_\perp^2}{2} + |h_{xy}|^2 + h_z^2 
	\pm \tfrac{t_\perp}{2}\sqrt{t_\perp^2 + 4|h_{xy}|^2}},
	\label{eq:E_k_valence}
\end{align}
where $E^c_\pm$ denote the two conduction bands, and $E^v_\pm$ represent the two valence bands of the bilayer system. The Halden mass is therefore defined by $2 m_H=E^c-E^v$.

To investigate the influence of circularly polarized light on the bilayer Haldane model, 
we employ the Floquet formalism, which provides a natural framework to incorporate 
time-periodic perturbations from an external electromagnetic field into the system's Hamiltonian. 
When the lattice is irradiated with circularly polarized light, the vector potential takes the form
\begin{equation}
    \mathbf{A}(t) = A_0 \left(\gamma \sin \omega_0 t, \, \cos \omega_0 t \right),
\end{equation}
where $\gamma=\pm 1$ distinguishes right- and left-handed polarizations, 
$A_0 = E_0/\omega_0$ with $E_0$ the electric-field amplitude, 
and $\omega_0$ the driving frequency of the external light.

The resulting time-dependent Hamiltonian is expressed as
\begin{equation}
    \hat{H}_{{\gamma,\omega}}(t) = \hat{H}(K) + \hat{V}(t),
\end{equation}
where $\hat{H}(K)$ is the unperturbed Hamiltonian (see Eq.~\ref{eq:ham_kspace}), and the interaction with 
the external field is captured by
\begin{equation}
    \hat{V}(t) = \frac{e v_{\rm F}}{\hbar} \Big[ \tau A_x(t)\,\sigma_x + A_y(t)\,\sigma_y \Big].
\end{equation}
where $v_{\rm F}\sim 10^{6}$m/s represents the Fermi velocity of Dirac Fermions in monolayer graphene and $\tau$ refers to the valley index.

In the high-frequency regime, characterized by $\omega_0 \gg e v_{\rm F} A_0/\hbar$, 
and under weak-field conditions $e v_{\rm F} A_0 \ll \hbar \omega_0$, 
Floquet theory allows us to construct an effective static Hamiltonian,
\begin{equation}
    \hat{H}_{\text{eff}} = \hat{H} + \frac{[\hat{H}_{-1},\hat{H}_{+1}]}{\hbar \omega_0} 
    + \mathcal{O}(\omega_0^{-2}),
    \label{effH}
\end{equation}
where $\hat{H}_{\pm 1}$ are the first-order Fourier harmonics of the time-dependent Hamiltonian, where the central sideband has the most significant effect,
\begin{equation}
    \hat{H}_{\pm n} = \frac{\omega_0}{2\pi} 
    \int_0^{2\pi/\omega_0} dt\, \hat{H}(t) e^{\pm i n \omega_0 t}.
\end{equation}
and we neglect the higher order $H_{n}$ for $n>1$.

In the limit $(eaA_0/\hbar)^2 \ll 1$, the Floquet correction introduces a light-induced 
mass term of the form
\begin{equation}
    \lambda_{\omega} = \frac{[\hat{H}_{-1}, \hat{H}_{+1}]}{\hbar \omega_0} 
    = \gamma \frac{(e v_{\rm F} A_0)^2}{\hbar \omega_0},
    \label{omega}
\end{equation}
which explicitly depends on the polarization $\gamma$ and the field strength. The resulting effective $h_z(\mathbf{k}, \phi_p)$ component of the Hamiltonian described in Eq. (\ref{eq:ham_kspace}) is modified to $h_z(\mathbf{k}, \phi_p)+\lambda_{\omega}$, where $\lambda_{\omega}$ is given in Eq.(\ref{omega}). The value of $A_0$ is usually around $0.08$ nm$^{-1}$.

The Floquet-induced coupling fundamentally reshapes the quasiparticle spectra described by Eqs.~\eqref{eq:E_k_conduction} and \eqref{eq:E_k_valence} through the opening of dynamical band gaps. Therefore, the band gap is controlled by topology and the Floquet-induced coupling.

\subsection{Low-energy expansion at the semi-Dirac merging point}

We derive the approximated low-energy effective Hamiltonian close to the Dirac-merging point in order to see the band touching at the crucial anisotropy. The Bloch Hamiltonian for the anisotropic honeycomb lattice with nearest-neighbor hoppings $t$ and $t'$ can be expressed in the sublattice basis as

\begin{equation}
H(\mathbf{k})=
\begin{pmatrix}
0 & f(\mathbf{k}) \\
f^*(\mathbf{k}) & 0
\end{pmatrix},
\end{equation}

where the structure factor is

\begin{equation}
f(\mathbf{k}) =
t' e^{i k_y a_0}
+ t e^{-i\left(\frac{\sqrt{3}}{2}k_x a_0+\frac{1}{2}k_y a_0\right)}
+ t e^{i\left(\frac{\sqrt{3}}{2}k_x a_0-\frac{1}{2}k_y a_0\right)} .
\end{equation}

As we will explain in the following, Dirac points merge when the anisotropy reaches the critical value at $t'=2t$.

At this point, the two Dirac cones collide at the $M$ point of the Brillouin zone, $\mathbf{k}_M=(0,\pi/a_0)$. We therefore expand the Hamiltonian around this momentum by defining small deviations $q_x = k_x$ and $q_y = k_y - \pi$. 

Substituting $k_y=\pi+q_y$ and expanding the structure factor to leading order in $q_x$ and $q_y$, we get

\begin{equation}
f(\mathbf{q}) \approx \hbar v_{\rm F} q_x + i \hbar \alpha q_y^2 ,
\end{equation}

where the coefficients are $\hbar v_{\rm F} = \sqrt{3}t/a_0$ and $\hbar \alpha = \frac{t}{2}/a_0^2$.

Therefore, the effective low-energy Hamiltonian near the merging point can be written as

\begin{equation}
H(\mathbf{q}) =
\hbar v_{\rm F} q_x \sigma_x +
\hbar \alpha q_y^2 \sigma_y ,
\end{equation}

which corresponds to the well-known semi-Dirac Hamiltonian. The resulting energy dispersion becomes
$E(\mathbf{q}) =
\pm \hbar \sqrt{ v_{\rm F}^2 q_x^2 + \alpha^2 q_y^4 }$.

This spectrum represents a semi-Dirac point that results from the merger of two Dirac cones; it is linear along the $q_x$ direction but quadratic along $q_y$. The low-energy Hamiltonian for a bilayer graphene receives an extra contribution when the Haldane and Floquet-induced mass terms are present in a bilayer structure.

\begin{equation}
H_{\mathrm{eff}}(\mathbf{q}) =
-\frac{t^2}{t_{\perp}}[Re (g^2(q)) \sigma_x+Im (g^2(q)) \sigma_y]+
(m_H + \lambda_\omega) \sigma_z.
\label{eq:lowenergy}
\end{equation}
where $g(q)=1+\sqrt{3}a_0q_x+ia_0q_y$. 
The gap closes when the eigenenergy, $E=m_{{\mathrm{eff}}} \pm \frac{t^2}{t_{\perp}}|g(q)|^2$ vanishes, marking the topological phase transition associated with the redistribution of Berry curvature near the merging point.
Here $m_{\mathrm{eff}}=m_H + \lambda_\omega$ and therefore, the Floquet mass is competing with the intrinsic Haldane mass.
Expanding for a small momentum gives
\begin{equation}
    E= m_{\mathrm{eff}}\pm\frac{t^2}{t_{\perp}}(1+2\sqrt{3}a_0q_x+a_0^2 q_y^2)
\end{equation}
and the above expression is demonstrating that interlayer coupling renormalizes the semi-Dirac dispersion but preserves the anisotropic linear-quadratic character near the merging point.

\section{Spectral properties}\label{sec:bandstructure}

In this section, we analyze the evolution of the Floquet-engineered band structure as the system is continuously tuned between the Dirac and semi-Dirac regimes. This interpolation is achieved by varying the strength of the Floquet-induced mass term, $\lambda_{\omega}$, in conjunction with the interlayer coupling $t_{\perp}$ and the NNN hopping amplitude $t_2$. Representative band dispersions are presented for selected parameter sets to highlight the underlying spectral reconstruction.

We first consider the case $t_{\perp}=0.5t$ and $t_2=0$, as shown in Fig.~\ref{f1}. The resulting spectrum comprises four well-separated energy bands. For clarity, we label the two conduction bands as band-c1 (upper) and band-c2 (lower), and the valence sector as band-v2 (upper) and band-v1 (lower). In the absence of the Haldane flux, the system preserves inversion and time-reversal symmetries, and the low-energy physics is governed by massless Dirac fermions. Consequently, band-c2 and band-v2 exhibit gap closing and reopening at the Fermi level at the K and K$^\prime$ points of the Brillouin zone as $\lambda_{\omega}$ is varied (see Figs.~\ref{f1}(a)--\ref{f1}(f)). These band touchings correspond to symmetry-protected Dirac points, signaling a Floquet-controlled band-gap engineering mechanism.

As the hopping anisotropy is enhanced, the Dirac nodes migrate along high-symmetry directions in momentum space and progressively approach each other. Upon reaching a critical value, they merge at the $\mathbf{M}$ point, giving rise to a semi-Dirac dispersion characterized by linear energy-momentum scaling along one direction and quadratic scaling along the orthogonal direction. In this regime, the Floquet-induced term $\lambda_{\omega}$ continues to act as an effective mass, driving successive gap opening and closing transitions even beyond the Dirac merging point, as illustrated in Figs.~\ref{f1}(g)--\ref{f1}(l).

The spectral properties are qualitatively modified once the Haldane flux is introduced ($t_2 \neq 0$), as shown in Fig.~\ref{f2}. The NNN hopping explicitly breaks time-reversal symmetry and generates a topological mass term, leading to a robust gap opening at the K and ${K}^\prime$ valleys for small hopping anisotropy. While the overall gap evolution with $\lambda_{\omega}$ remains analogous to the $t_2=0$ case in the Dirac regime, a distinct behavior emerges near the semi-Dirac transition. Specifically, for $t_1 \gtrsim 1.5t$, the conduction and valence bands close and reopen precisely at the semi-Dirac point as the Floquet driving frequency is varied. This gap-closing mechanism closely resembles that observed in monolayer graphene at the Dirac-to-semi-Dirac transition, where the band touching marks a critical point separating distinct topological phases \cite{MondalS2021}.
  
We examine the regime of stronger interlayer hybridization together with the impact of light polarization. The results appear in the Appendix.

Based on our analysis, the Dirac points shift toward each other with increasing hopping anisotropy and ultimately merge at the $\mathbf{M}$ point, signaling the transition to the semi-Dirac regime. Simultaneously, the enhanced interlayer coupling continues to increase the energetic separation between the conduction bands (band-c1 and band-c2) as well as between the valence bands (band-v1 and band-v2). These results highlight the combined influence of interlayer hybridization and light helicity in enabling tunable, valley-selective Floquet band engineering. From an effective low-energy perspective, the observed polarization-dependent gap modulation can be interpreted in terms of a valley-contrasting Floquet mass term. Circularly polarized light generates an effective time-reversal-symmetry-breaking mass whose sign depends both on the helicity of the driving field and on the valley index. As a result, the induced Floquet mass acquires opposite signs at the K and K' points for a fixed light polarization, leading to valley-selective gap closing and reopening at distinct critical values of $\lambda_{\omega}$. Reversing the light helicity inverts the sign of the effective mass, thereby exchanging the roles of the two valleys. This behavior is a hallmark of Floquet-driven valley control and provides a direct mechanism for engineering valley-contrasting band inversions and topological phase transitions in the strong interlayer coupling regime.

\section{Chern number and phase diagram}\label{sec:phase_diagram}

In this section, we investigate the topological character of the bilayer system by computing the Chern numbers as functions of the Haldane fluxes threading the two layers. Since time-reversal symmetry is explicitly broken, the bands acquire finite Chern numbers, which are obtained from the Berry curvature integrated over the Brillouin zone (BZ) \cite{ThoulessD1998,AvronJE1988},
\begin{eqnarray}\label{eq:chern_number}
	C & = & \frac{1}{2\pi} \int\int_{\mathrm{BZ}}\!\!\Omega(k_x, k_y)\, \mathrm{d}k_x \mathrm{d}k_y ,
\end{eqnarray}
where $\Omega(k_x, k_y)$ is the $z$-component of the Berry curvature \cite{LiuCXZhang2016}, defined as
\begin{eqnarray}\label{eq:berry_curv}
	\Omega(k_x, k_y) = -2i\,\mathrm{Im}\!\left[\left\langle \frac{\partial\psi(k_x, k_y)}{\partial k_x} \,\Bigg|\, \frac{\partial\psi(k_x, k_y)}{\partial k_y}\right\rangle\right],
\end{eqnarray}
with $\psi(k_x, k_y)$ denoting the periodic part of the Bloch wavefunction corresponding to the Hamiltonian in Eq.~\eqref{eq:ham_kspace}. For numerical purposes, the Wilson-loop approach might be used to ensure the gauge-independent result. The larger Chern number yields the larger total Berry flux, the stronger anomalous velocity contribution, and the enhanced transverse response.  

We present the Berry curvature ($\Omega$) distributions for the valence bands $v_1$ and $v_2$ in Figs.~\ref{f7} and \ref{f8}, respectively. In both cases, the Berry curvature is strongly localized near the band extrema in momentum space. Upon introducing hopping anisotropy in the range $t < t_1 < 2t$, the band extrema move closer to one another, resulting in a corresponding migration and distortion of the regions carrying large Berry curvature. This evolution is clearly visible through the progressive deformation of the honeycomb-like patterns in Fig.~\ref{f7}(\ref{f7b})-Fig.~\ref{f7}(\ref{f7c}) for band $v_1$ and similarly in Fig.~\ref{f8}(\ref{f8b})-Fig.~\ref{f8}(\ref{f8c}) for band $v_2$.
At a Haldane flux $\phi_l=\phi_u=\pi/2$, the Chern number associated with band $v_2$ equals $-2$ for $t_1<2t$, whereas band $v_1$ remains topologically trivial with a vanishing Chern number. Nevertheless, finite Berry curvature persists for both bands over the entire range of $t_1$, reflecting the fact that time-reversal symmetry is explicitly broken throughout. For $t_1>2t$ (e.g., $t_1=2.2t$), the Berry curvature of band $v_2$ undergoes a clear sign reversal, as shown in the Fig.~\ref{f8}(\ref{f8c}). This demonstrates that a nonvanishing Berry curvature distribution does not necessarily imply a finite Chern number.
Finally, the Floquet-induced term $\lambda_{\omega}$ further reshapes the Berry curvature profiles while keeping all other parameters fixed. This effect is manifested as a systematic modification of the curvature intensity and distribution when moving from left to right within each row for a given value of $t_1$, highlighting the role of Floquet driving as an additional control knob for Berry-curvature engineering.

Moreover, because of the coupled Haldane flux and circularly polarized drive, the curvature in the nontrivial phases is substantially concentrated at the K and K′ valleys, indicating broken time-reversal symmetry. A single dominant hotspot produces a net Berry flux of $\pm 2π$ in the $C =\pm 1$ phase. On the other hand, the $C=\pm 2$ regime shows several constructive hotspots of uniform sign, suggesting two co-propagating chiral edge modes and simultaneous band inversion in two hybridized Dirac sectors as well as an accumulated Berry flux of $\pm 4π$.

The curvature becomes more anisotropic and moves in the direction of the $\mathbf{M}$ point as the hopping anisotropy gets closer to the Dirac-merging condition, reflecting the underlying semi-Dirac dispersion. Topological charge transfer across bands is indicated by the curvature sharpening around the gap-closing momentum at the critical point. The curvature distribution is globally inverted when the light helicity is reversed, indicating that the Floquet-induced mass is the main topological phase control parameter.

Figure~\ref{f9} summarizes the resulting topological phase diagrams for different values of the hopping anisotropy $t_1$ at fixed interlayer coupling $t_{\perp}=0.5t$, focusing first on the valence band $v_1$. The red and blue regions correspond to Chern insulating phases with Chern numbers $C=+1$ and $C=-1$, respectively, while the white areas denote a topologically trivial phase with $C=0$. As evident from Fig.~\ref{f10}, the Chern insulating regions attain their maximum extent at $t_1=t$, corresponding to the Dirac limit. In this regime, the Floquet-induced term $\lambda_{\omega}$ plays a decisive role in stabilizing the topological phases, with $\lambda_{\omega}=0.025t$ yielding the largest Chern regions for fixed $t_1=t$.
As $t_1$ is increased, the size of these regions, commonly referred to as ``Chern lobes''—progressively shrinks and nearly vanishes by $t_1\simeq1.9t$, as shown in the last column of Fig.~\ref{f10}. In this regime, increasing $\lambda_{\omega}$ enhances the Chern insulating phase space, demonstrating that Floquet driving can partially compensate for the suppression of topology induced by hopping anisotropy. At the semi-Dirac point $t_1=2t$, all Chern numbers vanish as a consequence of the band touching between band-$v_2$ and band-$c_2$, despite band-$v_1$ remaining gapped. For $t_1>2t$, a finite gap reopens; however, the system remains topologically trivial.

A richer phase structure emerges when considering the valence band $v_2$, which lies closer to the Fermi level, as shown in Fig.~\ref{f10}. In this case, additional Chern insulating phases with higher topological charge $C=\pm2$ appear, indicated by cyan ($+2$) and green ($-2$) regions where topology is collapsing at semi-Dirac merging. Consequently, phases with $C=\pm2$ and $C=\pm1$ coexist within the same phase diagram, depending on the values of the Haldane fluxes $\phi_l$ and $\phi_u$. Similar to band-$v_1$, the corresponding Chern lobes contract with increasing $t_1$ and collapse entirely at the semi-Dirac point $t_1=2t$, beyond which the system remains gapped but topologically trivial. The phase diagrams for the conduction bands $c_1$ and $c_2$ mirror those of $v_1$ and $v_2$, respectively, with opposite Chern numbers due to particle–hole symmetry. In all cases, the Floquet-induced coupling $\lambda_{\omega}$ plays an analogous role, controlling the stability and extent of the Chern insulating phases across the entire spectrum. Notice that higher Chern bands are extremely important in interacting systems; namely, flat $C=2$ bands can host fractional Chern insulators, and an analog of higher Landau levels, which are highly sought in moir\'e systems.

We can conclude that the system's bilayer structure, which essentially offers two topological Dirac sectors per valley, is the source of the appearance of $C =\pm 2$ phases. Two co-propagating chiral edge modes and a twofold Chern number result from the constructive addition of the Berry curvature contributions of both sectors undergoing band inversion with the same mass sign. This multi-channel structure is stabilized by interlayer hybridization, which also enables Floquet driving to concurrently invert many bands, creating higher-Chern insulating phases that are not present in the monolayer limit.

\section{Hall conductivity} \label{sec:hall_conductivity}

We now turn to the anomalous Hall response of the system and compute the Hall conductivity as a function of the Fermi energy $E_{\rm F}$. The starting point is the Berry curvature, given in Eq.~\eqref{eq:berry_curv}, which enters directly into the expression for the anomalous Hall conductivity \cite{XiaoDChang2007},
\begin{equation}\label{eq:Hall_cond}
	\sigma_{xy} = \frac{\sigma_0}{2\pi} \sum_{j} \int \frac{\mathrm{d}k_x \,\mathrm{d}k_y}{(2\pi)^2} \,
	f\!\left(E^j_{k_x, k_y}\right)\,\Omega(k_x, k_y) ,
\end{equation}
where $E^j_{k_x, k_y}$ denotes the energy of the band indexed by $j$, and 
\begin{equation}
    f(E^j_{k_x, k_y})=\Big[1+\exp\!\left\{\frac{E^j_{k_x, k_y}-E_{\rm F}}{K_BT}\right\}\Big]^{-1}
\end{equation}
is the Fermi–Dirac distribution at temperature $T$. The scale of the conductivity is set by $\sigma_0=e^2/h$. In what follows, we evaluate $\sigma_{xy}$ numerically at zero temperature ($T=0$) for various values of the hopping asymmetry $t_1$, as shown in Fig.~\ref{fhc}. Figure~\ref{fhc} presents the anomalous Hall conductivity $\sigma_{xy}$ as a function of the Fermi energy $E_{\rm F}$ for several values of the hopping anisotropy $t_1$, with the left and right panels corresponding to interlayer couplings $t_{\perp}=0.5t$ and $t_{\perp}=1.5t$, respectively. The remaining parameters are fixed at $t_2=0.1t$, while $t_1$ is varied as $t_1=t$, $1.5t$, $1.8t$, and $2.2t$. Although, $E_{\rm F}\ll t$ and greater than the energy gap for real materials, for the sake of completeness, we also demonstrate the results for $E_{\rm F}\ge t$. The effect of Floquet driving is examined by considering three representative amplitudes of the Floquet-induced term, $\lambda_{\omega}=0.025t$ (dotted line), $\lambda_{\omega}=0.1t$ (dashed line), and $\lambda_{\omega}=0.225t$ (solid line). When the Fermi energy resides inside the bulk band gap, the Hall conductivity exhibits a quantized plateau at $\sigma_{xy}=2\sigma_0$, signaling the emergence of a Chern insulating phase. The width of this plateau is directly governed by the size of the bulk gap, in full agreement with the corresponding band dispersions shown in Fig.~\ref{f1}. A systematic dependence on the Floquet amplitude is clearly observed: for all values of $t_1$, the plateau is widest for $\lambda_{\omega}=0.025t$, becomes narrower for $\lambda_{\omega}=0.225t$, and is smallest for $\lambda_{\omega}=0.1t$. This behavior demonstrates that the Floquet term $\lambda_{\omega}$ serves as an efficient knob for tuning the bulk gap and, consequently, the energy window over which the Hall conductivity remains quantized. As soon as $E_{\rm F}$ enters either the conduction or valence bands, the quantization breaks down and $\sigma_{xy}$ decreases continuously. This reduction arises from the partial occupation of bulk states, as the Berry curvature contribution in Eq.~\eqref{eq:Hall_cond} is then integrated over an incomplete set of filled bands. Increasing the hopping anisotropy $t_1$ leads to a progressive narrowing of the quantized plateau, reflecting the gradual reduction of the gap between band-c2 and band-v2. At the semi-Dirac transition ($t_1 \gtrsim 1.8t$), the gap closes completely, resulting in the disappearance of the plateau and a concomitant sign change in the Hall conductivity. For $t_1>2t$, a gap reopens; however, $\sigma_{xy}$ acquires the opposite sign, indicating a topological phase transition. These features are fully consistent with the Chern number analysis. Quantized Hall plateaus appear only when the system resides in a Chern insulating phase ($t_1<2t$), with the quantization value $\sigma_{xy}=2\sigma_0$ directly reflecting a Chern number $C=2$ and the presence of two chiral edge modes. Beyond the semi-Dirac regime, the reversal of the Hall response signals a change in the underlying topological invariant.

For comparison, we also display $\sigma_{xy}$ for the weaker interlayer coupling $t_{\perp}=0.5t$ in the left panel of Fig.~\ref{fhc}. In this case, the quantized plateaus are noticeably narrower than those obtained for $t_{\perp}=1.5t$, consistent with the reduced bulk gaps observed in Figs.~\ref{f1} and \ref{f2}. Nevertheless, the qualitative behavior remains unchanged: the plateau width decreases with increasing $t_1$, vanishes near the semi-Dirac point, and reemerges with an opposite sign for $t_1>2t$, confirming the robustness of the Floquet-driven topological response across different interlayer coupling regimes.

As we show in Eq. (\ref{eq:lowenergy}), the low-energy physics near the Dirac-merging point in the driven system without the hybrization term $t_{\perp}$ can be described by the effective Floquet semi-Dirac Hamiltonian.
The Berry curvature associated with the semi-Dirac sector can be expressed as

\begin{equation}
\Omega(\mathbf{q}) =
-\frac{\alpha v_{\rm F} m_{\mathrm{eff}} q_y}
{\left(v_{\rm F}^2 q_x^2 + \alpha^2 q_y^4 + m_{\mathrm{eff}}^2\right)^{3/2}},
\end{equation}

which shows that the Berry curvature is proportional to the effective mass. Consequently, the Chern number and the anomalous Hall conductivity satisfy

\begin{equation}
\sigma_{xy} = C\frac{e^2}{h}, \qquad
C \propto \mathrm{sgn}(m_{\mathrm{eff}}).
\end{equation}
 
Reversing the light helicity flips $m_{\mathrm{eff}}$ and inverts the Berry curvature distribution over the Brillouin zone by changing the sign of the Floquet mass $\lambda_\omega$. This immediately results in the anomalous Hall conductivity's observed sign reversal.

\section{Conclusion}\label{sec:conclusion}

In summary, we have systematically investigated the topological properties of a band-engineered bilayer Haldane model driven by off-resonant circularly polarized light. By tuning the intralayer nearest-neighbor hopping anisotropy while keeping the interlayer coupling $t_{\perp}$ fixed, and by incorporating a Floquet-induced mass term $\lambda_{\omega}$, we demonstrated that the low-energy band extrema move continuously from the nonequivalent K and K' points toward the $\mathbf{M}$ point of the Brillouin zone. The merger at $t_1=2t$ defines the semi-Dirac limit, where the quasiparticle dispersion becomes linear along one momentum direction and quadratic along the other. 
A detailed analysis of the Chern numbers across the $(\phi_u,\phi_l)$ parameter space reveals a remarkably rich topological landscape. Higher Chern sectors with $C=\pm 2$ emerge exclusively in the valence band closest to the Fermi level (band-$v_2$), while both valence bands undergo sharp topological transitions characterized by abrupt changes in their Chern numbers. These include direct transitions such as $C=\pm 2 \rightarrow 0$ and $C=\pm 1 \rightarrow 0$, as well as multistep pathways involving sign reversals and intermediate phases, e.g., $+2 \rightarrow -2$ and $\pm 2 \rightarrow \pm 1 \rightarrow 0$. Each transition is accompanied by a gap-closing event in the bulk spectrum, signaling the emergence of critical semimetallic states at the phase boundaries.
The anomalous Hall conductivity provides an independent and consistent probe of these topological transitions. The quantized Hall plateau at $2\sigma_0$, associated with a Chern number $C=2$, progressively narrows as the hopping anisotropy $t_1$ is increased and undergoes a complete sign reversal at the semi-Dirac point $t_1=2t$. Beyond this critical value, the conductivity reemerges with opposite sign, in direct correspondence with the inversion of the underlying topological invariant. Taken together, our results establish that the bilayer Haldane model hosts a genuine Floquet-driven topological phase transition at the semi-Dirac limit, analogous in spirit to the monolayer case but with substantially enriched phenomenology. The bilayer geometry supports multiple chiral edge channels, enhanced quantized Hall responses, and a more intricate evolution of Chern numbers across the phase diagram. The combined effects of band engineering, Floquet control, and interlayer hybridization thus position bilayer Haldane systems as a versatile platform for realizing and manipulating higher-Chern-number phases and light-tunable topological states. Finally, the predicted phases may be found in experiments by using Hall conductivity measurements and spectroscopic probes such as ARPES or optical measurements, which can determine Floquet band modifications and topological edge states.

\section{acknowledgments}
I.K. and G.X. acknowledge the financial support from the NSFC under grant No. 12174346.

\appendix

\section{Details of band dispersion}\label{detaild}
\begin{figure*}[h!]
	\centering \includegraphics[width=15.90cm]{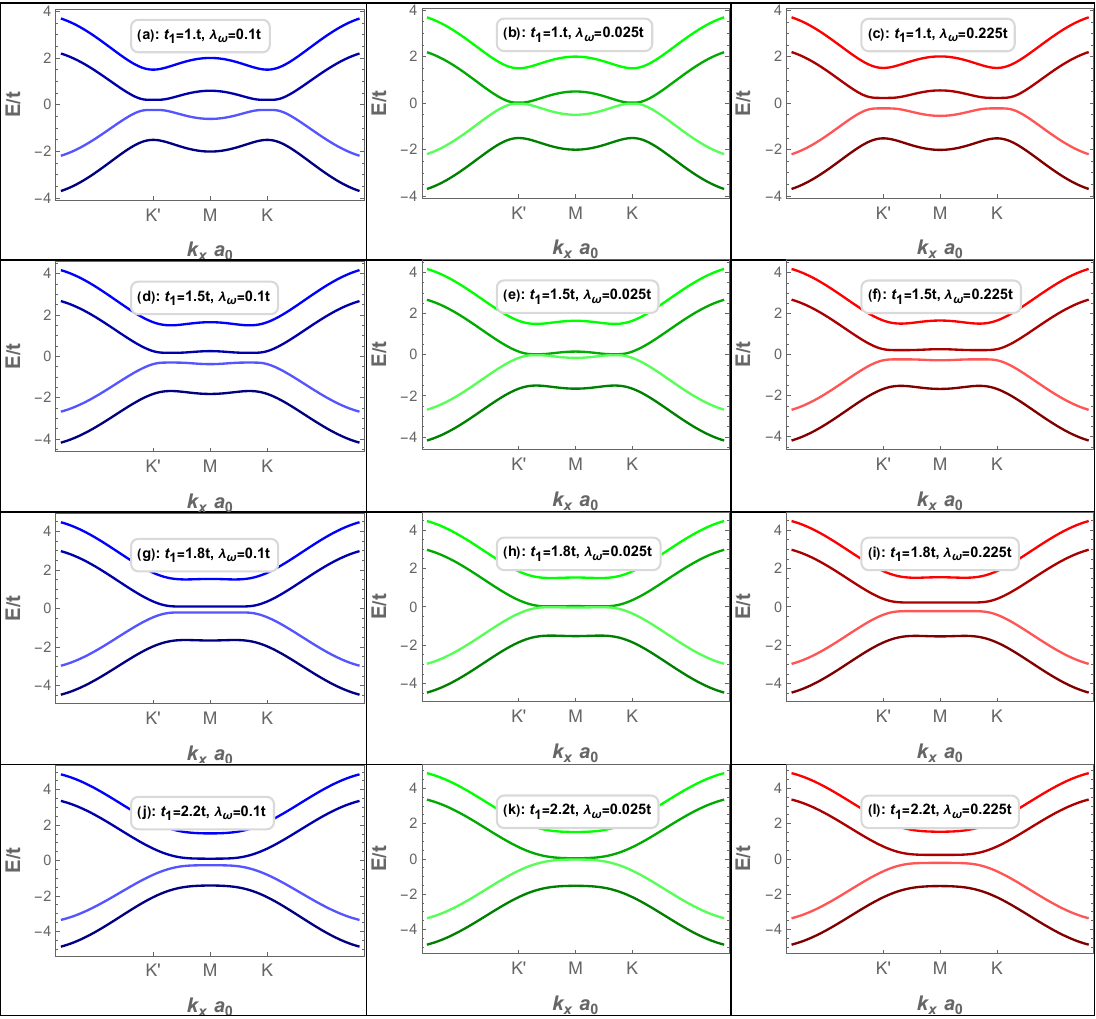}
   \caption{Band structures along the $k_x a_0$ direction at fixed $k_y a_0 = 2\pi/3$. Panels (a)--(l) correspond to finite next-nearest-neighbor hopping $t_2 = 0.0t$ with 
$t_1 = t$, $1.5t$, $1.8t$, and $2.2t$, respectively. For comparison, panels (e)--(h) show the dispersions in the absence of $t_2$ for the same sequence of $t_1$ values. The remaining parameters are fixed at $t_\perp = 1.5t$, and $\phi_l = \phi_u = \pi/2$.}
		\label{f3}
\end{figure*}
\begin{figure*}[h!]
	\centering \includegraphics[width=15.90cm]{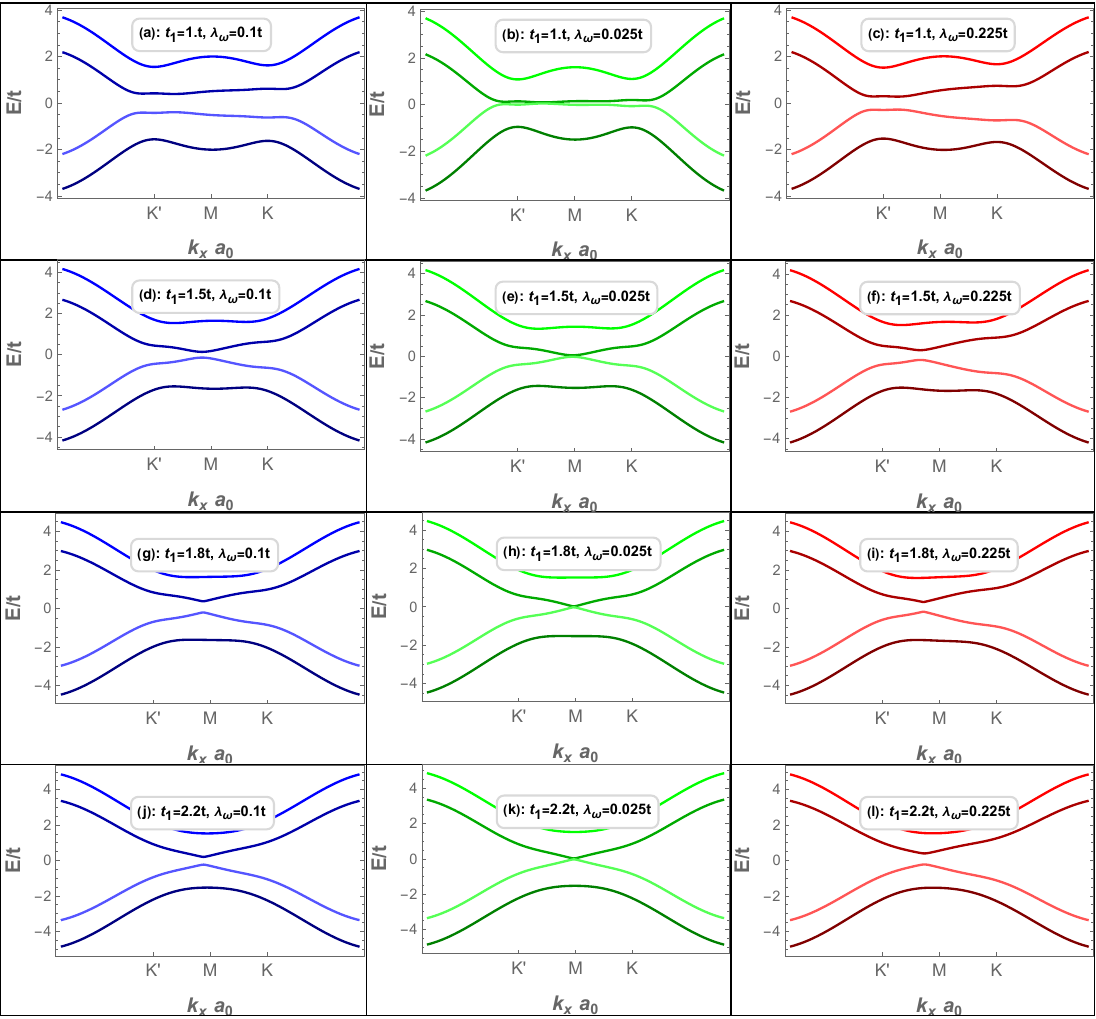}
   \caption{Band structures along the $k_x a_0$ direction at fixed $k_y a_0 = 2\pi/3$. 
Panels (a)--(l) correspond to finite next-nearest-neighbor hopping $t_2 = 0.1t$ with 
$t_1 = t$, $1.5t$, $1.8t$, and $2.2t$, respectively. For comparison, panels (e)--(h) 
show the dispersions in the presence of $t_2$, the same sequence of 
$t_1$ values. The remaining parameters are fixed at $t_\perp = 1.5t$,  
and $\phi_l = \phi_u = \pi/2$.}
		\label{f4}
\end{figure*}

\begin{figure*}[h!]
	\centering \includegraphics[width=15.90cm]{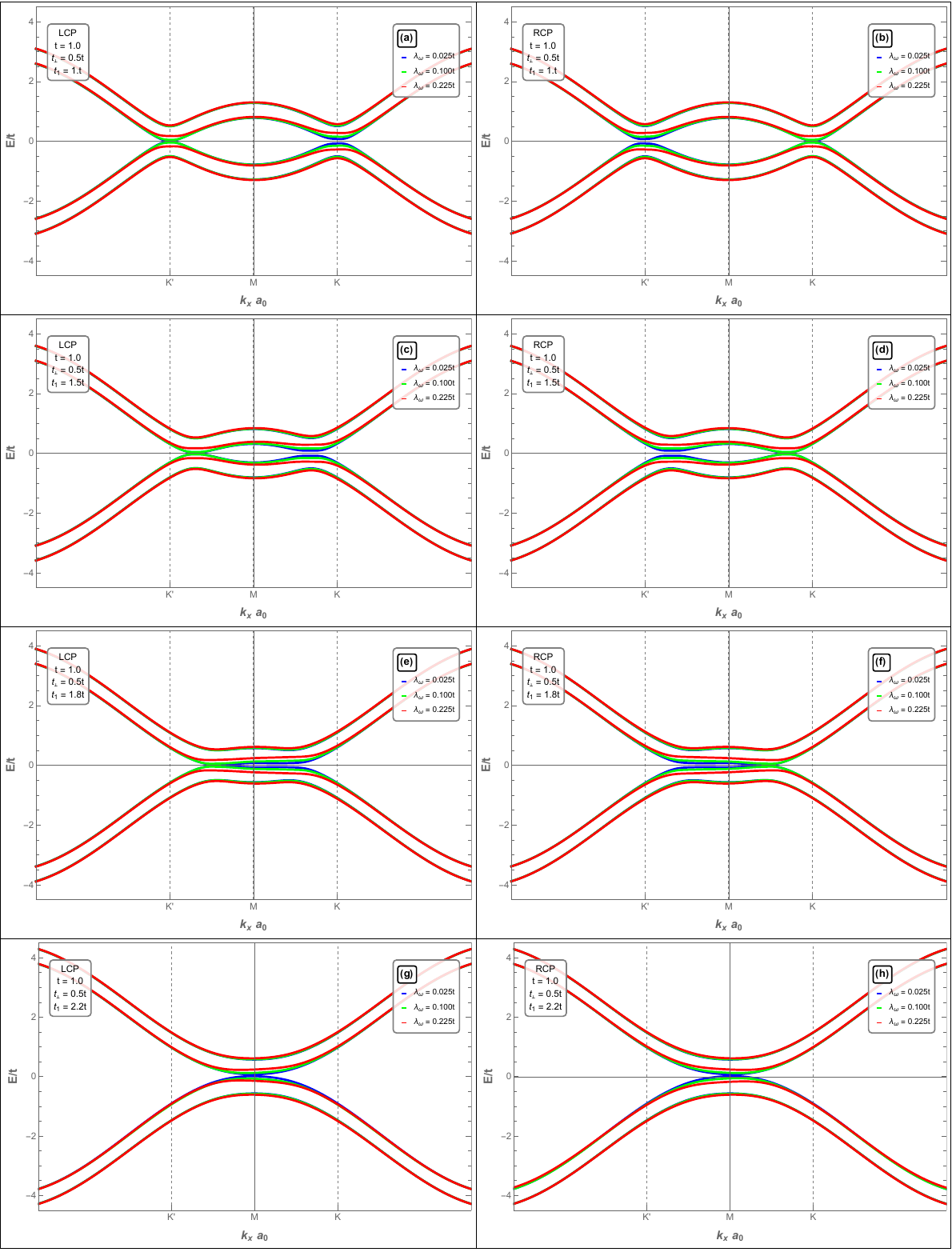}
   \caption{Band structures along the $k_x a_0$ direction at fixed $k_y a_0 = 2\pi/3$. 
Panels (a)--(l) correspond to finite next-nearest-neighbor hopping $t_2 = 0.0t$ with 
$t_1 = t$, $1.5t$, $1.8t$, and $2.2t$, respectively. For comparison, panels (e)--(h) 
show the dispersions in the absence of $t_2$ for the same sequence of 
$t_1$ values for the LCP and RCP light. The remaining parameters are fixed at $t_\perp = 0.5t$,  
and $\phi_l = \phi_u = \pi/2$.}
		\label{f5}
\end{figure*}
\begin{figure*}[h!]
	\centering \includegraphics[width=15.90cm]{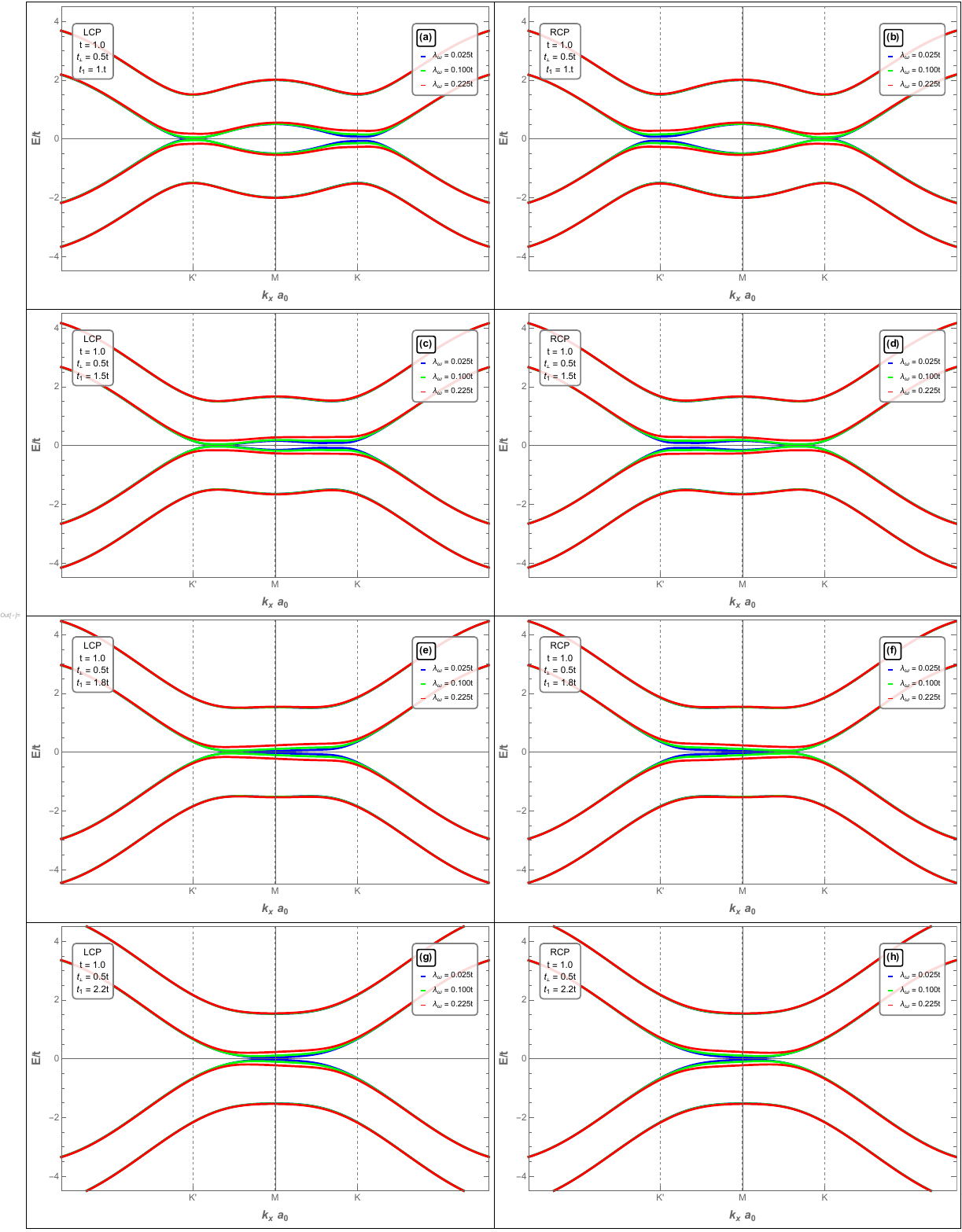}
   \caption{Band structures along the $k_x a_0$ direction at fixed $k_y a_0 = 2\pi/3$. 
Panels (a)--(d) correspond to finite next-nearest-neighbor hopping $t_2 = 0.1t$ with 
$t_1 = t$, $1.5t$, $1.8t$, and $2.2t$, respectively. For comparison, panels (e)--(h) 
show the dispersions in the presence of $t_2$, the same sequence of 
$t_1$ values for the LCP and RCP light. The remaining parameters are fixed at $t_\perp = 1.5t$,  
and $\phi_l = \phi_u = \pi/2$.}
		\label{f6}
\end{figure*}
In this section, we provide excess information regarding the band dispersion in terms of their physical parameters by combining bilayer Haldane, anisotropic hopping, and Floquet drive system. Importantly, we demonstrate that the semi-Dirac merging controls higher-Chern phases. We examine the regime of stronger interlayer hybridization, focusing on $t_{\perp}=1.5t$, as illustrated in Figs.~\ref{f3} and \ref{f4}. As anticipated from the analytical expressions for the quasiparticle energies in Eqs.~\eqref{eq:E_k_conduction} and \eqref{eq:E_k_valence}, an increase in $t_{\perp}$ significantly enhances the energetic separation between the two conduction bands (band-c1 and band-c2) as well as between the valence bands (band-v1 and band-v2). This band repulsion reflects the strengthened interlayer tunneling, which promotes pronounced hybridization between the layer-resolved electronic states. A notable consequence of the enhanced interlayer coupling is the qualitative modification of the low-energy dispersion near the band-touching points. In contrast to the linear Dirac-like spectra observed for $t_{\perp}=0.5t$, the dispersions of band-c2 and band-v2 now acquire a predominantly quadratic character. This crossover signals a substantial increase in the effective quasiparticle mass, indicating that the low-energy carriers evolve toward massive Dirac-like excitations as $t_{\perp}$ is further increased. At the same time, stronger interlayer coupling leads to a systematic reduction of the energy gap between band-c2 and band-v2. This gap suppression becomes increasingly pronounced as the system approaches the semi-Dirac regime at larger values of the hopping anisotropy $t_1$. The observed behavior highlights the delicate interplay between interlayer tunneling and band topology, demonstrating that it $t_{\perp}$ serves as an efficient control parameter for engineering both the dispersion anisotropy and the magnitude of the low-energy spectral gap. Importantly, the Floquet-induced term $\lambda_{\omega}$ continues to play the same qualitative role as in the $t_{\perp}=0.5t$ case: the gap opening and closing transitions occur at identical Floquet conditions, indicating that the critical values of $\lambda_{\omega}$ governing these spectral reconstructions remain unchanged by the enhanced interlayer coupling.

We next investigate the role of light polarization in shaping the Floquet-engineered band structure, focusing on the strong interlayer coupling regime $t_{\perp}=1.5t$. The corresponding band dispersions for right-circularly polarized (RCP) and left-circularly polarized (LCP) light are presented in Figs.~\ref{f5} and \ref{f6}, respectively. A pronounced polarization-dependent response is observed at the nonequivalent peaks K and K' valleys of the Brillouin zone, reflecting the valley-selective nature of circularly polarized driving. For RCP illumination, the gap-closing behavior differs between the two valleys: at the K point, the gap is closed by the Floquet-induced term $\lambda_{\omega}=0.1t$, as indicated by the green band, whereas at the K' point the gap closure occurs at a smaller Floquet amplitude, $\lambda_{\omega}=0.025t$, corresponding to the blue band. This asymmetric response between the valleys signals a polarization-controlled breaking of valley degeneracy. Upon reversing the helicity of the light, i.e., for LCP driving, the roles of the K and K' points are interchanged, demonstrating a clear helicity-dependent valley selectivity of the Floquet-induced band inversions.

Apart from this polarization-driven valley contrast, the remaining spectral features remain qualitatively consistent with those discussed in Figs.~\ref{f1}--\ref{f4}.

\end{document}